\documentclass[aps,prc,twocolumn,superscriptaddress]{revtex4-1}
\usepackage{graphicx}
\usepackage{multirow}
\usepackage{natbib}
\usepackage{color}
\usepackage{ulem}
\usepackage{amsmath}

\begin{document}


\title{Fission barrier, damping of shell correction and neutron emission in the fission of A$\sim$200}


\author{K.~Mahata}
\email{kmahata@barc.gov.in}
\affiliation{Nuclear Physics Division, Bhabha Atomic Research Centre, Mumbai 400 085, INDIA}
\affiliation{Homi Bhabha National Institute, Anushakti Nagar, Mumbai 400 094, INDIA}
\author{S.~Kailas}
\email{Raja Ramanna Fellow}
\affiliation{Nuclear Physics Division, Bhabha Atomic Research Centre, Mumbai 400 085, INDIA}
\affiliation{UM-DAE Centre for Excellence in Basic Sciences, Mumbai 400 098, INDIA}
\author{S.~S.~Kapoor}
\email{INSA Honorary Scientist}
\affiliation{Nuclear Physics Division, Bhabha Atomic Research Centre, Mumbai 400 085, INDIA}


\date{\today}

\begin{abstract}
Decay of $^{210}$Po compound nucleus formed in light and heavy-ion induced fusion reactions has been analyzed simultaneously using a consistent prescription for fission barrier and nuclear level density incorporating shell correction and its damping with excitation energy.  Good description of all the excitation functions  have been achieved with a fission barrier of 21.9 $\pm$ 0.2 MeV. For this barrier height, the predicted statistical pre-fission neutrons  in heavy-ion fusion-fission are much smaller than the experimental values, implying the presence of dynamical neutrons due to dissipation even at these low excitation energies ($\sim$ 50~MeV) in the mass region A $\sim$ 200.  When only heavy-ion induced fission excitation functions and the pre-fission neutron multiplicities are included in the fits, the deduced best fit fission barrier depends on the assumed fission delay time during which dynamical neutrons can be emitted.  
A fission delay of (0.8 $\pm$ 0.1 )$\times 10^{-19}$ s has been estimated corresponding to the above fission barrier height assuming that the entire excess neutrons over and above the statistical model predictions are due to the dynamics.  The present observation has implication on the study of fission time scale/ nuclear viscosity using neutron emission as a probe.

\end{abstract}


\maketitle

The fission process involves most drastic rearrangements in nuclei, where both statistical and dynamical features, governed by the delicate interplay between the macroscopic (liquid drop) aspects  and the quantal (shell) effects, are exhibited.
One of the key questions in nuclear fission is: what is the maximum energy along the fission path (barrier height)~\cite{Moller09}? 
The fission barrier has contributions from
the macroscopic liquid drop part as well as from  the microscopic shell effects. Accurate knowledge of the fission barrier 
height is  vital not only to understand the heavy-ion induced fusion-fission 
dynamics and  predictions concerning super-heavy nuclei, but also other areas, 
such as stellar nucleosynthesis and nuclear energy applications. The status of charged particle  induced fission reactions has been reviewed recently~\cite{Kailas14}

Experimental determination of the fission barrier height in mass A $\sim$ 200 continues to be a challenging problem. 
In this mass region, the fission barrier heights are much higher than the neutron separation energies and experimental cross sections around the fission barrier, being extremely low, are often not available. 
Large ground state shell corrections around the Z=82, N=126 brings in  additional parameters in the investigation of fission in mass A $\sim$ 200 region.   

\begin{figure}[tb]
\begin{center}
\includegraphics[width=0.37\textwidth]{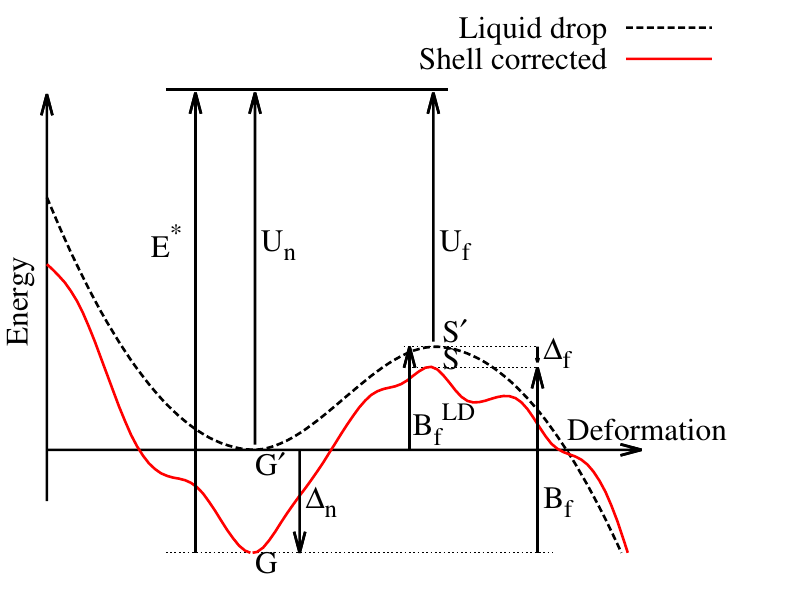}
\caption{(Color online) A schematic representation of the potential energy as a function of deformation in mass A $\sim$ 200. For large E*, shell effects are washed out and the available thermal energy (U) is measured with respect to liquid drop surface.}
\label{bf}
\end{center}
\vskip-2ex
\end{figure}

As shown in Fig.~\ref{bf}, the mass of a nucleus gets lowered from the liquid drop (LD) ground state due to negative shell correction energy ($\Delta_n$). Knowledge of the shell correction at the saddle deformation ($\Delta_f$) for A$\sim$200 is obscure and most of the analyses assume $\Delta_f = 0$.  The shell correction reduces with increasing excitation energy and washes out at  excitation energy of around 30-40~MeV. A clear manifestation of this is observed in the energy dependence of the nuclear level density.  For the ground state shape, the nuclear 
level density of a closed shell nucleus shows the same energy dependence as  nuclei away from the shell closure at high excitation energies ($E^*>$ 30 to 40  MeV) if  the thermal energy U is measured from the liquid drop ground state i.e. back-shifted by $-\Delta_n$ ($U = E^* + \Delta_n$)~\cite{Ramamurthy70}. At intermediate energies the dependence is accounted   by an energy dependent level density parameter $a(E^*)$, approaching asymptotically to the liquid drop value ($\tilde{a}$)~\cite{Ramamurthy70,Ignatyuk75}, which is equivalent to an energy dependent back-shift. It may be interpreted as if a part of the excitation energy is spent to melt the shell correction partially or fully and the rest appears as thermal energy.

According to the statistical model of compound
nucleus decay, all possible decays are equally likely and the decay probabilities are governed by the relative density of levels (phase space)~\cite{Stokstad85}. 
Level density for the ground state shape $\rho_n(E^*)$ at the excitation energy E$^*$ can be  calculated as~\cite{Ramamurthy70,Ignatyuk75}
\begin{eqnarray}
[\ln\rho_n(E^*)]^2 &\sim& 2\tilde{a_n}\left[1 + \frac{\Delta_n}{E^*}(1 - e^{-\eta E^*})\right]E^*\nonumber \\
               &\sim& 2\tilde{a_n}[\underbrace{E^* + \Delta_n(1 - e^{-\eta E^*})}_{U_n(E^*,\Delta_n)}]
\end{eqnarray}
For large E$^*$, shell effects are washed out as $\eta E^*$ is large and the asymptotic value of $\rho_n(E^*)$ is given by [$\ln\rho_n(E^*)]^2 = 2\tilde{a_n}[E^* + \Delta_n]$. Similarly, level density for the saddle shape $\rho_f(E^*)$ corresponding to the fission barrier height B$_f$ can be  calculated as
{\small{\begin{eqnarray}
[\ln\rho_f(E^*)]^2 &\sim& 2\tilde{a_f}\left[1 + \frac{\Delta_f}{E^*-B_f}(1 - e^{-\eta (E^*-B_f)})\right](E^*-B_f)\nonumber \\
               &\sim& 2\tilde{a_f}[\underbrace{E^* -B_f + \Delta_f(1 - e^{-\eta (E^*-B_f)})}_{U_f(E^*,\Delta_f)}]
\end{eqnarray}}}
For the saddle point shape also as excitation energy ($E^*-B_f$) reaches around 30-40 MeV the shell effects are washed out and nuclear level density can be calculated by measuring excitation energy with respect to the liquid drop surface. Using $B_f = B_f^{LD} - \Delta_n + \Delta_f$, the asymptotic value of $\rho_f(E^*)$ can be expressed as  [$\ln\rho_f(E^*)]^2 = 2\tilde{a_f}[E^* + \Delta_n - B_f^{LD}]$. Therefore the saddle point level density at high excitation energy does not depend  on any shell correction which may be present at the saddle point. The asymptotic LD value of the level density parameter at the saddle point ($\tilde{a_f}$ ) may be different from that for the ground state configuration ($\tilde{a_n}$), because of the differences in the nuclear shapes.  In the above discussion, we have ignored the angular momentum dependence of the level density for simplicity and assumed same damping constant ($\eta$) for $\Delta_n$ and $\Delta_f$.

Several statistical model analyses, particularly for the  decay of $^{210}$Po, have been presented in the past. Considering that shell corrections are washed out at high excitation energies, Charity et al.~\cite{Charity86} have calculated $U_n$ using liquid drop (LD) mass (G$^\prime$) and used $U_f = U_n - B_{LD}(J)$ in the statistical model  analysis of the measured fission and evaporation residue (ER) excitation functions for $^{18}$O+$^{192}$Os system.
Moretto et al.~\cite{Moretto95} have also considered that shell corrections are washed out at excitation energies above 20 MeV and  have taken $U_n = E^* + \Delta_n$, $U_f = E^* - B_{LD}(J) + \Delta_n$, in the analysis of fission excitation function for $\alpha$ + $^{206}$Pb system. 

As can be seen from the Eq.~2 as well as from the Fig.~1, neither the available thermal energy nor the level density parameter at the saddle deformation should  depend on the shell correction at the ground state ($\Delta_n$) and its damping with energy. However, this might not have been followed in many of the analyses reported in literature~\cite{DArrigo94,Shrivastava99,Schmitt14}. 
In a statistical model analysis~\cite{DArrigo94} of the fission excitation functions for $p$+$^{209}$Bi, $\alpha$ + $^{206}$Pb, $^{12}$C + $^{198}$Pt and $^{18}$O+$^{192}$Os systems, populating $^{210}$Po, smooth damping of shell corrections were considered by having  energy dependent level density parameters. 
However, the same shell correction ($\Delta_n$) was used to calculate the energy dependence of $a_n$ and $a_f$ with different damping functions.  
In an another study~\cite{Shrivastava99} of the fission and ER excitation functions for $^{12}$C + $^{198}$Pt system, the fission barrier was considered to have only macroscopic part and same value of shell correction ($\Delta_n$) was used to calculate excitation energy dependence  of $a_n$ and $a_f$. This amounts to lowering  of the whole LD surface ($G^\prime S^\prime$) by an  equal amount ($\Delta_n$) and gradual damping of the shell correction with excitation energy. In a recent 4D Langevin calculation~\cite{Schmitt14} for $^{12}$C + $^{198}$Pt system,  the E$^*$ has been used as internal excitation (thermal) energy instead of E$^*+ \Delta_n$ and the potential energy surface was considered to have only macroscopic part. This also amounts to lowering of the whole liquid drop surface (the equilibrium as well as the saddle deformation) by the same amount ($\Delta_n$). As the value of $\Delta_n$ is large (-10.6 MeV) for $^{210}$Po, lowering of the whole potential energy surface is expected to have significant effect on some of the calculated quantities.

As pointed out earlier, at the high excitation energies involved in the heavy-ion fusion-fission reaction, the shell effects  are washed out. Hence, information on the shell corrections at the saddle point and the fission barrier height  can not be obtained by fitting available data on heavy-ion fusion-fission excitation functions at hight excitation energies alone. The measured fission excitation functions in heavy-ion induced reactions could be explained by several pairs of correlated values of 
($B_f$, $\tilde {a_f}/\tilde {a_n}$), but the predicted pre-fission neutron multiplicities ($\nu_{pre}$) were found to be sensitive to this correlated variation~\cite{Vigdor80,Ward83,Mahata03,Mahata06}.
For this reason, in our earlier work~\cite{Mahata06} simultaneous analysis of the measured fission excitation function and pre-fission neutron multiplicities ($\nu_{pre}$) was thought to be a way to fix these parameters. In this earlier work, the pre-fission neutron multiplicities were corrected for the emission of dynamical neutrons corresponding to an assumed fission delay of 30$\times 10^{-21}$s, taken from the literature~\cite{Hinde89}. Further, in the earlier work, $U_n$ and $U_f$ were calculated on the basis of Eq. 1 and 2, respectively.   
Best fit to the fission and evaporation residue  excitation functions along with the estimated statistical part of pre-fission neutron multiplicity data for $^{12}$C+$^{198}$Pt system required a value of $B_f(0)$ = 13.4 MeV, implying a significant shell correction at the saddle point~\cite{Mahata06}. 

\begin{figure}[tb]
\begin{center}
\includegraphics[trim = 0mm 4mm 0mm 0mm, clip,width=0.4\textwidth]{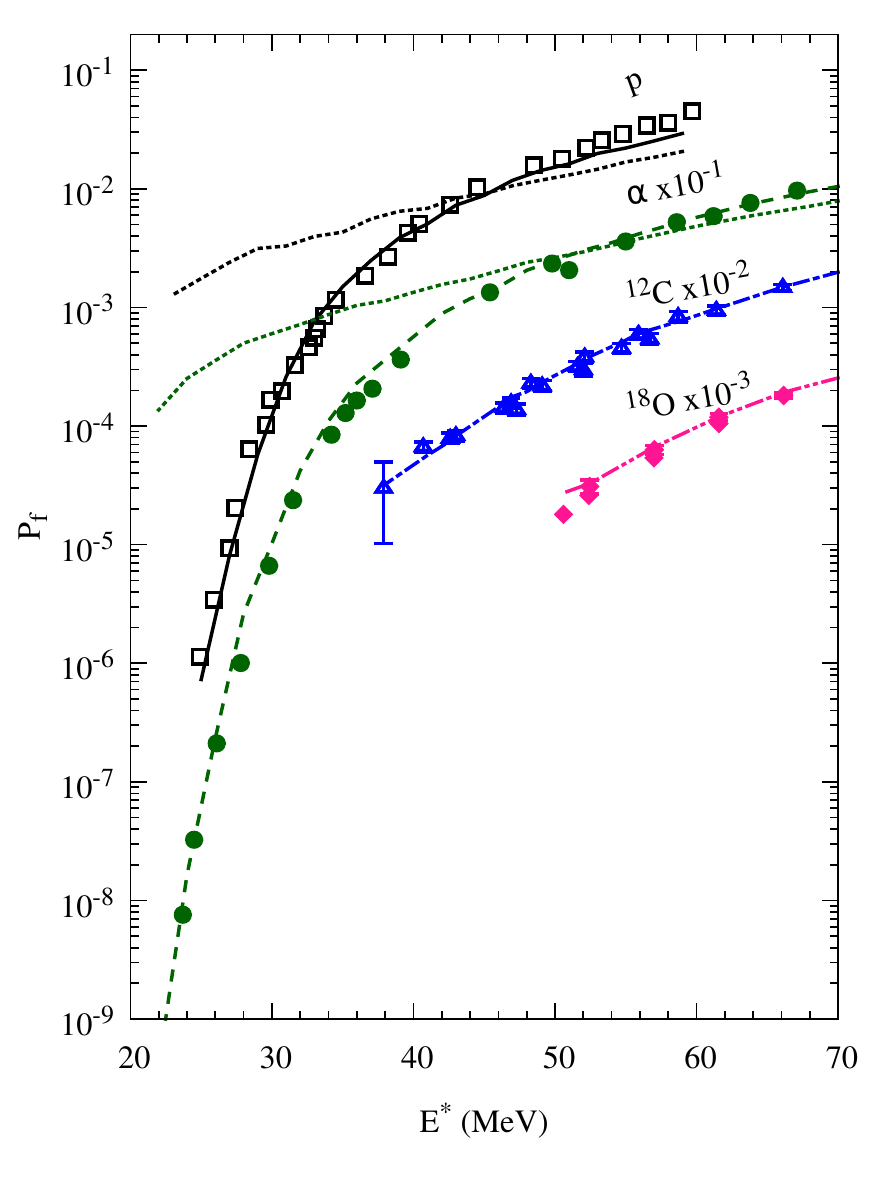}
\caption{(Color online) Experimental fission probabilities of $^{210}$Po are compared with the statistical model calculations. The (black) continuous, (green) dashed, (blue) dot-dashed and the (pink) dot-dot-dashed  lines are the statistical model predictions with B$_f$(0) = 21.9 MeV, $\tilde {a_f}/\tilde {a_n}$ = 1.027 for fusion of p,  $\alpha$, $^{12}$C and $^{18}$O projectiles, respectively. The black and green dotted lines are the predictions of the statistical model with B$_f$ = 13.4 MeV~\cite{Mahata06} for fusion of $p$ and $\alpha$ particles, respectively. 
}
\label{pf}
\end{center}
\vskip-2ex
\end{figure}

In the present work, we have analyzed the partial evaporation residue excitation functions, fission excitation functions 
for $p$, $\alpha$, $^{12}$C and $^{18}$O induced fusion reactions forming the same compound nucleus $^{210}$Po, without including $\nu_{pre}$ data.
The experimental fission probabilities, shown in Fig.~\ref{pf}, are calculated as  P$_f$ = $\sigma_{fis}/\sigma_{fus}$ using the experimental fission cross-sections ($\sigma_{fis}$) for $p$~\cite{Khodai66, Zhukova77, Ignatyuk84}, $\alpha$~\cite{Moretto95}, $^{12}$C~\cite{Shrivastava99, Plicht83} and $^{18}$O~\cite{Charity86, Plicht83} induced fusion reactions. In case of $p$ induced reaction  the sum of $xn$ evaporation~\cite{Exfor} and fission cross-sections have been taken as the fusion cross-section ($\sigma_{fus}$). For the $\alpha$ induced reaction, the fusion cross-section have not been measured experimentally and have been taken from  the Bass model~\cite{Bass77}. The Bass model gives a good description of the measured fusion excitation function for $\alpha$+$^{209}$Bi system~\cite{Penionzhkevich02}. Effect of pre-equilibrium particle emission in the case of $p$ and $\alpha$ induced reactions in the present analysis is estimated to be insignificant~\cite{Mahata14a}. Experimental fusion cross-sections for $^{12}$C and $^{18}$O have been taken from the literature~\cite{Shrivastava99,Charity86}.



The analysis has been carried out using the code PACE~\cite{Gavron80} with the modified prescriptions for fission barrier and level density
parameter~\cite{Mahata06}. The  excitation energy of the compound nucleus
is taken as $E^* = E_{c.m.} + Q - E_{rot}(J) -\delta_p$, where $E_{c.m.}$, Q, $E_{rot}$ and $\delta_p$ are the energy in the centre-of-mass system, Q-value for fusion, rotational energy and pairing energy, respectively. The Q-value and particle separation energies for subsequent decays are calculated using the experimental mass~\cite{Wang12}. The rotational energy  $E_{rot}(J)$ is taken from Ref.~\cite{Sierk86}. 
The damping of shell correction is taken into account by having the thermal energy at equilibrium deformation as $U_n  = E^* + \Delta_n(1 - e^{-\eta E^*})$. As shown in Eq.~1, this is equivalent to  Ignatyuk prescription~\cite{Ignatyuk75}. The value of level density parameter ($\tilde {a_n}$) is taken as A/9. Shell corrections at the ground state ($\Delta_n$) are taken from Ref.~\cite{Myers94}.  The fission barrier is  expressed as 
$B_f(J) = B_f^{RFRM}(J) - \Delta_n + \Delta_f$.
The angular momentum dependent macroscopic part of the fission barrier ($B_f^{RFRM}(J)$) is taken from the
Rotating Finite Range Model~\cite{Sierk86}. The thermal excitation energy at the saddle deformation is calculated as $U_f = E^* - B_f(J) + \Delta_f(1 - e^{-\eta (E^*-B_f)})$. In the earlier analysis~\cite{Mahata06}, simultaneous fit to  
fission and ER cross sections along with $\nu_{pre}$ values for $^{12}$C+$^{198}$Pt system required B$_f$(0) to be 13.4 MeV with $\Delta_f$ = 0.76$\times \Delta_n$. However, the value of fission probabilities  (P$_f$) for $p$ and $\alpha$ induced
fission at lower excitation energies, which are more sensitive to the value of B$_f$ and less sensitive to $\tilde {a_f}/\tilde {a_n}$~\cite{Mahata14a},  are grossly over-predicted with the use of such a small value of B$_f$ (large value of $\Delta_f$) as shown by the (black and green) dotted lines in Fig.~\ref{pf} for fusion of $p$ and $\alpha$ particles. Hence, the value of $\Delta_f$ and $\tilde {a_f}/\tilde {a_n}$ have been varied to fit the experimental excitation functions. An excitation energy dependent shell damping factor $\eta = 0.054 + 0.002\times E^*$
for the equilibrium configuration is found to reproduce the shape of the measured fission excitation function better than a constant damping factor~\cite{Mahata14a}. Best fit to the $p$ and $\alpha$ induced
fission data results in $\Delta_f = 0.4 \pm 0.3$ MeV and 0.7 $\pm$ 0.3 MeV, respectively. The corresponding values of $\tilde {a_f}/\tilde {a_n}$ are 1.036 and 1.018 for $p$ and $\alpha$ induced
fission, respectively. As stated earlier, the fission excitation functions for $^{12}$C and $^{18}$O projectiles are not sensitive to the correlated variation of $\Delta_f$ and $\tilde {a_f}/\tilde {a_n}$. It should be noted here that the uncertainty in the macroscopic part of the barrier as well as in the ground state shell correction has not been considered in deducing the above values of $\Delta_f$.

\begin{figure}[tb]
\begin{center}
\includegraphics[trim = 0mm 4mm 0mm 0mm, clip,width=0.38\textwidth]{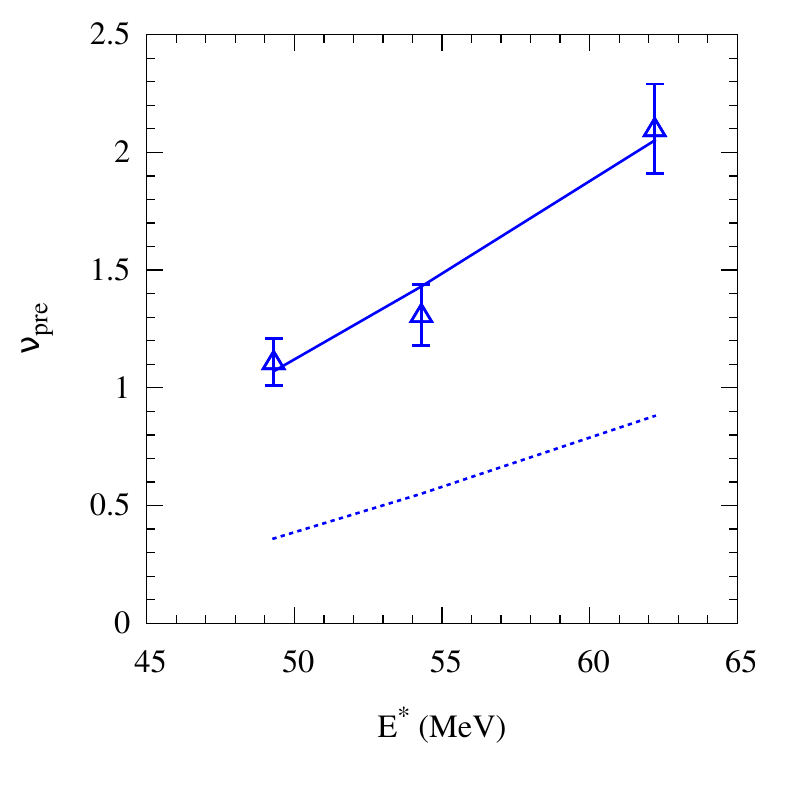}
\caption{(Color online) Experimental pre-fission neutron multiplicity data (open triangle)~\cite{Golda13} are compared with the statistical model calculations (blue, dotted line) with B$_f$(0) = 21.9 MeV for fusion of  $^{12}$C projectile. 
The (blue) continuous is obtained by adding the estimated dynamical neutrons corresponding to a fission delay of 0.8$\times10^{-19}$ to the statistical model calculation (see text).
}
\label{npre}
\end{center}
\end{figure}

A value of 21.9 $\pm$ 0.2 MeV for the fission
barrier, which is found to reproduce the fission and $xn$ excitation functions for all the systems populating $^{210}$Po, is in good agreement with the prediction (22.1 MeV) of the macroscopic-microscopic finite-range liquid-drop model~\cite{Moller09}.   As shown in Fig.~\ref{npre}, the statistical model calculation with this value of 
fission barrier predicts the statistical part of the $\nu_{pre}$ values which are much smaller than the experimental $\nu_{pre}$ value for fusion of $^{12}$C with $^{198}$Pt~\cite{Golda13}.

\begin{figure}[tb]
\begin{center}
\includegraphics[trim = 0mm 3mm 0mm 0mm, clip,width=0.38\textwidth]{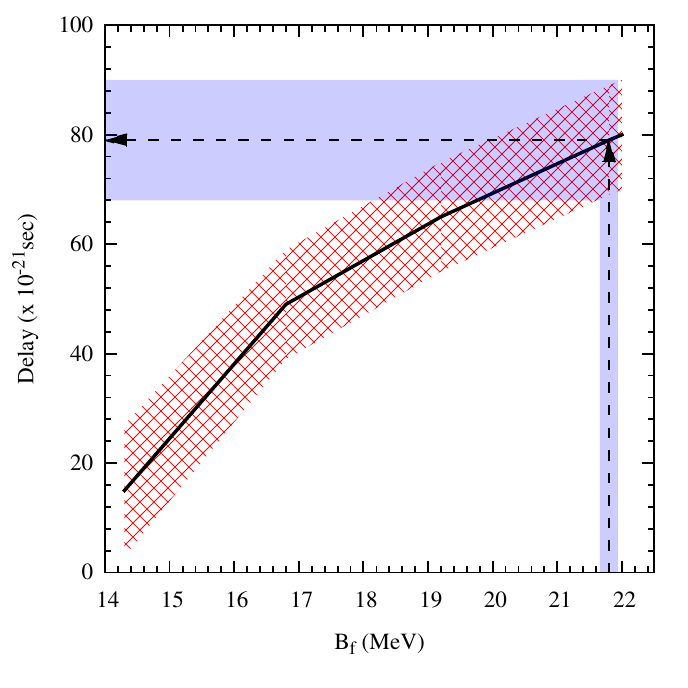}
\caption{(Color online) Correlation between the fission barrier height and the dynamical delay required to account for the excess neutrons as compared to the prediction of the statistical model. The hatched region corresponds to the uncertainty in the measured pre-fission neutron multiplicity data. The shaded region indicates the uncertainty in the fission barrier obtained from the analysis of all the fission excitation functions and the corresponding dynamical delay.}
\label{delay}
\end{center}
\end{figure}

From the above analysis, it is evident that while the fission barrier for $^{210}$Po is 21.9$\pm$0.2 MeV, and there are no significant shell corrections at the saddle point,
there is an excess emission of  neutrons as compared to the statistical model predictions even at excitation energy as low as 50 MeV. In the literature,  excess neutron emission as compared to the statistical model predictions has been observed due to the fission delay at higher excitation energies~\cite{Hinde89}. We have estimated fission delay assuming that these excess neutrons arise entirely due to dynamical emission using a Monte Carlo procedure. The mean lifetime of neutron emission ($\tau_n$) at each decay step is estimated from the neutron decay width ($\Gamma_n$) from the statistical model.  The time distribution of neutron emission is obtained by multiplying $\tau_n$ with the negative logarithm of a random
number, chosen in the interval between 0 and 1~\cite{Gavron87,Rossner89}. The contribution due to dynamical emission is taken as the ratio of the number of neutrons emitted within the dynamical delay time to the total number of cascades. In the present work, fission delay has been estimated under the assumption of dynamical emission taking place only at the equilibrium deformation. 
Emission at other deformations will be less probable due to lesser excitation energy available and a recent dynamical calculation~\cite{Nadtochy14} also suggests that the fissioning nuclei in this mass region spends most of its time in the equilibrium and  saddle region.

If both heavy-ion fusion-fission excitation function and prefission neutron data are included in the fit, the best fit fission barrier height depends on the assumed statistical component of the experimental prefission neutrons. Further,  the division of pre-fission neutrons into statistical and dynamical contributions depends on the value of fission delay during which dynamical neutrons can be emitted. Fig.~\ref{delay} shows the correlation between the best fit fission barrier height and the required fission delay assuming emission from equilibrium deformation.  On the basis  that dissipation may not be important at low energies (E$^* <$ 60 MeV) for this mass region~\cite{Back99,Thoennessen93}, in out earlier work~\cite{Mahata06}, a correction to $\nu_{pre}$ was made assuming dynamical delay of  0.3$\times10^{-19}$s, which resulted in much smaller fission barrier implying a large shell correction at the saddle point.  A dynamical delay of (0.8 $\pm$ 0.1)$\times10^{-19}$s is required for emission at equilibrium deformation corresponding to the fission barrier of 21.9 MeV to fit all the fission excitation functions.   As shown in Fig.~\ref{npre}, dynamical contributions corresponding to a delay of 0.8$\times10^{-19}$ s in addition to the statistical contributions reproduce the measured pre-fission neutron multiplicities well.

As discussed in Ref.~\cite{Hinde89}, the calculated neutron mean life and hance the extracted fission delay also depends on the value of level density parameter. In the above calculation the value of $\tilde{a}_n$ is assumed to be A/9, which gave a better reproduction of the experimental partial $xn$ excitation functions~\cite{Mahata06} and is also within the acceptable range of A/8 to A/9.5, determined from the measured neutron spectrum for a nearby compound nucleus ($^{208}$Pb)~\cite{Rout13}. 
Significant  near scission emission (0.1 - 0.3 neutrons/fission) has been reported in spontaneous and induced fission at low energies of actinide nuclei in the literature~\cite{Halpern71}.   If such non-statistical near scission neutron emission at the instant of neck breaking is also present in case of fission of lighter nuclei such as $^{210}$Po, the dynamical component will reduce leading to a fission delay  smaller than 0.8$\times10^{-19}$ s.

In summary, the decay of $^{210}$Po formed in fusion of $p$, $\alpha$, $^{12}$C and $^{18}$O projectiles has been analyzed simultaneously,  using a consistent prescription for fission barrier and level density allowing continuous damping of shell corrections. 
The low energy light-ion induced fission excitation functions are found to be crucial to determine the height of the fission barrier  accurately. 
The best fit fission barrier  of $^{210}Po$ is found to be B$_f$ = 21.9$\pm$0.2 comprised mainly of a liquid drop component (B$_{LD}$=10.8~MeV) and a ground state shell correction ($\Delta_n$=-10.6~MeV) without significant shell correction at the saddle point. The statistical model calculation with the fission barrier of 21.9 MeV  substantially under predicts the experimental  pre-fission neutron multiplicities in heavy-ion fusion-fission reactions, indicating the presence of non-statistical neutron emission. In view of this, the pre-fission neutron multiplicity data should not be used to constrain the statistical model parameter as attempted earlier~\cite{Vigdor80,Ward83,Mahata06}. Assuming that the bulk of these excess neutrons may be dynamical neutrons, a significant fission delay is implied even at low excitation energies of about 50 MeV in this mass region. 
For the heavy-ion induced fusion-fission, a correlation has been found between the best fit fission barrier and the assumed fission delay during which  dynamical neutron can be emitted.
Dynamical calculation with the fission barrier obtained from the present analysis should provide more detailed knowledge of fission dynamics in this mass region.  Since the fission delay deduced in this work depends on the value of pre-fission neutrons, more measurements of pre-fission neutron multiplicity at low energies in this mass region will be desirable. 
The present result indicating a significant fission delay at lower excitation energies is of much relevance to the study of nuclear viscosity using neutron emission as a probe.

One of us (SK) acknowledges the support received under the DAE Ramanna fellowship. One of us (SSK) acknowledges the support from Indian National Science Academy.


\providecommand{\noopsort}[1]{}\providecommand{\singleletter}[1]{#1}%

\end{document}